\documentclass[aps,prb,a4paper,preprint,showpacs]{revtex4}
\usepackage{amsmath}
\usepackage{amsfonts}
\usepackage{amssymb}
\usepackage{slashbox}
\usepackage{graphicx}
\usepackage{amsbsy}

\begin{document}
\title{Resonance in the nonadiabatic quantum pumping of the time-dependent Josephson junction }
\author{Rui Zhu\renewcommand{\thefootnote}{*}\footnote{Corresponding author.
Electronic address:
rzhu@scut.edu.cn} and Mi Liu}
\address{Department of Physics, South China University of Technology,
Guangzhou 510641, People's Republic of China }

\begin{abstract}

In this work, we investigated the nonadiabatic transport properties of the one-dimensional time-dependent superconductor-normal metal-superconductor (SNS) Josephson junction biased by a current source and driven by a high-frequency-ac-gate-potential applied to the normal-metal layer. BCS superconductors are considered and treated with the time-dependent Bogoliubov-de Gennes equation. Using Floquet theory, we compute the transmission coefficients and the Wigner-Smith delay times as a function of the incident energy and find that they display resonances when one of the electron or hole Floquet wavevectors coincides with the bound quasiparticle state within the superconducting energy gap. The resonance varies with the phase difference between the two superconductors as a result of the bound quasiparticle level displacement. The supercurrent flowing through the SNS junction is dramatically enhanced by the resonances.

\end{abstract}

\pacs {72.10.-d, 74.25.F-, 74.45.+c}

\maketitle

\narrowtext

\section{Introduction}

The scattering process often involves interferences between different quantum paths, in which constructive interference corresponds to resonance and destructive interference to antiresonance of the transmission. Asymmetric antiresonance with a minimum followed by a maximum is called a Fano resonance\cite{FanoPR1961, MiroshnichenkoRMP2010}. There have been a great number of studies devoted to resonance and/or antiresonance in various
quantum processes, such as scattered by an Anderson impurity\cite{LuoPRL2004}, tunneling through a quantum dot\cite{JohnsonPRL2004, BulkaPRL2001, KobayashiPRB2004}, scattering from a donor impurity in an
electron waveguide\cite{TekmanPRB1993, ZhuJPCM2013}, transport in spin inversion devices\cite{CardosoEPL2008}, Mie scattering in plasmonic nanoparticles and metamaterials\cite{LukyanchukNMat2010}, and etc. In nonadiabatic quantum pumping, Floquet sidebands are formed by high-frequency oscillating
potentials. In the case of a time-dependent quantum well, Fano resonance occurs in the transmission spectrum\cite{WLiPRB1999, JHDaiEPJB2014, RZhuJAP2015, Emmanouilidou2001}, pumped shot noise\cite{JHDaiEPJB2014, RZhuJAP2015}, and Wigner-Smith delay times (WSDT)\cite{WLiPRB1999, Emmanouilidou2001}, when the energy or wavevector of one of the Floquet levels matches the quasibound level inside. To our knowledge, these resonant effects have not been discovered in the time-dependent superconductor-normal metal-superconductor (SNS) Josephson junction driven by a nonadiabatic electric potential applied to the normal region.

By means of the Josephson effect, supercurrents can flow through the SNS junction biased by a current source\cite{BardeenPRB1972}. Bound quasiparticle states exist in the normal region of the SNS junction, each consisting of equal probabilities of particle and hole states. The Josephson effect is related to the Andreev reflection, the latter of which is defined by reflection of a particle into a hole at a pair-potential boundary with no change of current\cite{Andreev}. It can be conjectured from previous studies that in nonadiabatic quantum pumping of the SNS junction, resonance is possible to occur in the supercurrent when one of the Floquet sidebands coincides with one of the bound quasiparticle states within the energy gap of the superconductor, which will be confirmed by the present theoretic work.

Our approach is based on various development in methodology and concept recently. In classical mechanics, the duration time of a scattering process can be defined by the energy derivation of the action. In the absence of a time operator in quantum mechanics, the quantum analog of the duration time can be defined as the energy derivative of the quantum mechanical phase shift $\tau  = \hbar {{d\phi } \mathord{\left/
 {\vphantom {{d\phi } {dE}}} \right.
 \kern-\nulldelimiterspace} {dE}}$ during the scattering process. In a multi-channel or dynamic scattering process, the distribution of the WSDT ${\tau _1}, \cdots ,{\tau _N}$ are the eigenvalues of the $N \times N$ Wigner-Smith matrix $Q(E)= - i \hbar {S^\dag }{{\partial S} \mathord{\left/
 {\vphantom {{\partial S} {\partial E}}} \right.
 \kern-\nulldelimiterspace} {\partial E}}$ with $S$ the scattering or Floquet scattering matrix\cite{Emmanouilidou2001, Schomerus2015} and $N = 4\left( {2{n_{{\rm{max}}}} + 1} \right)$ with $n_{\rm{max}}$ the maximal Floquet channel index in the present dynamic electron-hole system. The density of states (DOS) $\rho \left( E \right)$ is directly related to $Q(E)$ and the WSDT $\tau _n$ by
\begin{equation}
\rho \left( E \right) = {\left( {2\pi \hbar } \right)^{ - 1}}{\rm{Tr}}Q\left( E \right) = {\left( {2\pi \hbar } \right)^{ - 1}}\sum\limits_n {{\tau _n}} .
\label{DOS}
\end{equation}
The parametric conductance derivatives of a quantum dot was investigated by relating it to the distribution of the Wigner-Smith time-delay Matrix\cite{Brouwer1997}. And also the WSDT were investigated in the scattering by strong time-periodic driving fields\cite{Emmanouilidou2001, WLiPRB1999}.
Experiments can yield information on the distribution of the WSDT. The driving field can be realized by applying an ultrahigh intensity laser or a local ac-signal top gate. In the research of the charge turnstile based on the superconducting hybrid structure, it was found that quantization of the current is affected by Andreev reflection and Cooper pair-electron cotunneling\cite{Averin2008}. Nonadiabatic effects in braidings of Majorana fermions in topological superconductors were studied using the time-dependent Bogoliubov-de Gennes (BdG) equations\cite{MengCheng2011}. The time-dependent BdG equation can be used to describe time-evolution
of BCS superconductors under a parametrically time-dependent
Hamiltonian\cite{MengCheng2011, Foster2013}. In this paper, we consider the transport properties of the time-dependent SNS Josephson junction biased by a current source and driven by a high-frequency-ac-gate-potential applied to the normal-metal layer. We use the Floquet scattering matrix method to solve the time-dependent BdG equation and relate the Floquet scattering matrix with the WSDT and the DOS. With the general relation between the Josephson supercurrent and the quasiparticle excitation spectrum\cite{BeenakkerPRL1991} as well as the bound quasiparticle states and the DOS of the continuous spectrum, the supercurrent is calculated.

\section{Theoretic Formalism }

We consider the nonadiabatic pumping properties in the one-dimensional time-dependent SNS Josephson junction biased by a current source and driven by a high-frequency-ac-gate-potential applied to the normal region. The time-dependent electric potential has the form of $V_1 \cos \omega t$. Width of the normal region is $L$. The considered model is sketched in Fig. 1. The driving field can be realized by applying a local ac-signal top gate. Assuming the junction
is located in the $x$-direction, the time-dependent BdG equation can be expressed as\cite{MengCheng2011}
\begin{equation}
i\frac{d}{{dt}}\left[ {\begin{array}{*{20}{c}}
{u\left( {x,t} \right)}\\
{v\left( {x,t} \right)}
\end{array}} \right] = {H_{{\rm{BdG}}}}\left( t \right)\left[ {\begin{array}{*{20}{c}}
{u\left( {x,t} \right)}\\
{v\left( {x,t} \right)}
\end{array}} \right].
\label{TimeDependentBdGEquation}
\end{equation}
The time-dependent BdG Hamiltonian is
\begin{equation}
{H_{{\rm{BdG}}}}\left( t \right) = \left[ {\begin{array}{*{20}{c}}
{h\left( {x,t} \right)}&{\Delta \left( x \right)}\\
{{\Delta ^\dag }\left( x \right)}&{ - h\left( {x,t} \right)}
\end{array}} \right],
\end{equation}
with
\begin{equation}
h\left( {x,t} \right) =  - \frac{{{\hbar ^2}}}{{2m}}\frac{{{\partial ^2}}}{{\partial {x^2}}} - {E_F} + U\left( {x,t} \right),
\end{equation}
\begin{equation}
U\left( {x,t} \right) = \theta \left( x \right)\theta \left( {L - x} \right){V_1}\cos \left( {\omega t} \right),
\end{equation}
\begin{equation}
\Delta \left( x \right) = \left\{ {\begin{array}{*{20}{l}}
{{\Delta _0}{e^{{\phi  \mathord{\left/
 {\vphantom {\phi  2}} \right.
 \kern-\nulldelimiterspace} 2}}},}&{x \le 0,}\\
{0,}&{0 \le x \le L,}\\
{{\Delta _0}{e^{ - {\phi  \mathord{\left/
 {\vphantom {\phi  2}} \right.
 \kern-\nulldelimiterspace} 2}}},}&{x \ge L.}
\end{array}} \right.
\end{equation}
$\theta (x)$ is the step function.

In advance of the time-dependent treatment, we consider
the bound quasiparticle states in the normal region within the energy gap of the superconductor. The quasiparticle energy $\varepsilon $ is measured with respect to the Fermi energy $E_F$. The current-flux normalized eigenfunctions of the static BdG equations are\cite{BeenakkerPRL1991}
\begin{equation}
\begin{array}{*{20}{l}}
{{r_e}\psi _{{S1},e}^ -  + {r_h}\psi _{{S1},h}^ + ,}&{x \le 0,}\\
{a\psi _{N,e}^ +  + b\psi _{N,e}^ -  + c\psi _{N,h}^ +  + d\psi _{N,h}^ - ,}&{0 \le x \le L,}\\
{{t_e}\psi _{{S2},e}^ +  + {t_h}\psi _{{S2},h}^ - ,}&{x \ge L,}
\end{array}
\end{equation}
with
\begin{equation}
\begin{array}{l}
\psi _{N,e}^ \pm  = \left( {\begin{array}{*{20}{c}}
1\\
0
\end{array}} \right){\left( {{k_e}} \right)^{ - {1 \mathord{\left/
 {\vphantom {1 2}} \right.
 \kern-\nulldelimiterspace} 2}}}\exp \left( { \pm i{k_e}x} \right),\\
\psi _{N,h}^ \pm  = \left( {\begin{array}{*{20}{c}}
0\\
1
\end{array}} \right){\left( {{k_h}} \right)^{ - {1 \mathord{\left/
 {\vphantom {1 2}} \right.
 \kern-\nulldelimiterspace} 2}}}\exp \left( { \pm i{k_h}x} \right),\\
\psi _{{S1},e,h}^ \pm  = \left( {\begin{array}{*{20}{c}}
{{e^{i\eta 1_{e,h}/2}}}\\
{{e^{ - i{{\eta 1_{e,h}} \mathord{\left/
 {\vphantom {{\eta 1_{e,h}} 2}} \right.
 \kern-\nulldelimiterspace} 2}}}}
\end{array}} \right){\left( {2{q_{e,h}}} \right)^{ - {1 \mathord{\left/
 {\vphantom {1 2}} \right.
 \kern-\nulldelimiterspace} 2}}}{\left( {{{{\varepsilon ^2}} \mathord{\left/
 {\vphantom {{{\varepsilon ^2}} {\Delta _0^2}}} \right.
 \kern-\nulldelimiterspace} {\Delta _0^2}} - 1} \right)^{ - {1 \mathord{\left/
 {\vphantom {1 4}} \right.
 \kern-\nulldelimiterspace} 4}}}\exp \left( { \pm i{q_{e,h}}x} \right),
\end{array}
\end{equation}
while $\psi _{{S2},e,h}^ \pm $ has ${\eta 1_{e,h}}$ replaced by ${\eta 2_{e,h}}$ and $\exp \left( { \pm i{q_{e,h}}x} \right)$ replaced by $\exp \left[ { \pm i{q_{e,h}}\left( {x - L} \right)} \right]$. Here, ${k_{e,h}} = {\left( {{{2m} \mathord{\left/
 {\vphantom {{2m} {{\hbar ^2}}}} \right.
 \kern-\nulldelimiterspace} {{\hbar ^2}}}} \right)^{{1 \mathord{\left/
 {\vphantom {1 2}} \right.
 \kern-\nulldelimiterspace} 2}}}{\left( {{E_F} + {\sigma _{e,h}}\varepsilon } \right)^{{1 \mathord{\left/
 {\vphantom {1 2}} \right.
 \kern-\nulldelimiterspace} 2}}}$, ${q_{e,h}} = {\left( {{{2m} \mathord{\left/
 {\vphantom {{2m} {{\hbar ^2}}}} \right.
 \kern-\nulldelimiterspace} {{\hbar ^2}}}} \right)^{{1 \mathord{\left/
 {\vphantom {1 2}} \right.
 \kern-\nulldelimiterspace} 2}}}{\left[ {{E_F} + {\sigma _{e,h}}{{\left( {{\varepsilon ^2} - \Delta _0^2} \right)}^{{1 \mathord{\left/
 {\vphantom {1 2}} \right.
 \kern-\nulldelimiterspace} 2}}}} \right]^{{1 \mathord{\left/
 {\vphantom {1 2}} \right.
 \kern-\nulldelimiterspace} 2}}}$, $\eta {1_{e,h}} = {\phi  \mathord{\left/
 {\vphantom {\phi  2}} \right.
 \kern-\nulldelimiterspace} 2} + {\sigma _{e,h}}\arccos \left( {{\varepsilon  \mathord{\left/
 {\vphantom {\varepsilon  {{\Delta _0}}}} \right.
 \kern-\nulldelimiterspace} {{\Delta _0}}}} \right)$, $\eta {2_{e,h}} =  - {\phi  \mathord{\left/
 {\vphantom {\phi  2}} \right.
 \kern-\nulldelimiterspace} 2} + {\sigma _{e,h}}\arccos \left( {{\varepsilon  \mathord{\left/
 {\vphantom {\varepsilon  {{\Delta _0}}}} \right.
 \kern-\nulldelimiterspace} {{\Delta _0}}}} \right)$, ${\sigma _e} = 1$, and ${\sigma _h} = -1$. The square roots are to be taken such that ${\mathop{\rm Re}\nolimits} {q_{e,h}} \ge 0$, ${\mathop{\rm Im}\nolimits} {q_e} \ge 0$, and ${\mathop{\rm Im}\nolimits} {q_h} \le 0$. The function $\arccos t \in \left[ {0,{\pi  \mathord{\left/
 {\vphantom {\pi  2}} \right.
 \kern-\nulldelimiterspace} 2}} \right]$ for $0 \le t \le 1$, while $\arccos t =  - i\ln \left[ {t + {{\left( {{t^2} - 1} \right)}^{{1 \mathord{\left/
 {\vphantom {1 2}} \right.
 \kern-\nulldelimiterspace} 2}}}} \right]$ for $t \ge 1$. Solvability of the continuity equations of the wave functions and their derivatives at $x=0$ and $x=L$ gives rise to the secular equation. Roots of $\varepsilon $ of the secular equation are the bound quasiparticle energies $E_b$ measured from the Fermi energy $E_F$. These bound states consist of equal probabilities of particle and hole states. Numerical results of $E_b$ as functions of $\Delta _0$ and $\phi$
are shown in Fig. 2. For a small superconductor energy gap $\Delta _0$ relative to the Fermi energy $E_F$ and a small normal-region width $L$ considered here, only a single bound quasiparticle state is sustained in the normal region\cite{BardeenPRB1972}. The bound quasiparticle level is close to the gap surface and increases linearly with the gap width. It varies in an absolute trigonal function with the phase difference $\phi$ between the two superconductors. It approaches maximum at $\phi=0$ and $2\pi $ and approaches minimum at $\phi = \pi$. These results reproduce those of Bardeen and Johnson\cite{BardeenPRB1972} in 1972.

Now we use the Floquet scattering theory to solve the time-dependent BdG equation (\ref{TimeDependentBdGEquation}). Wave functions in the superconductor and normal regions can be written as
\begin{equation}
\psi \left( {x,t} \right) = \sum\limits_{n =  - \infty }^{ + \infty } {{e^{ - i{\varepsilon _n}t}}\left\{ {\begin{array}{*{20}{l}}
{{\psi _{S1,n}},}&{x \le 0,}\\
{{\psi _{N,n}},}&{0 \le x \le L,}\\
{{\psi _{S2,n}},}&{x \ge L,}
\end{array}} \right.}
\label{TimeDependentWaveFunction}
\end{equation}
with ${\varepsilon _n} = \varepsilon  + n\hbar \omega $, $\varepsilon$ an energy within the static continuous quasiparticle spectrum above the energy gap, $n$ an integer varying from $- \infty$ to $+ \infty$, and
\begin{equation}
\begin{array}{l}
{\psi _{S1,n}} = a_n^l\psi _{S1,e,n}^ +  + b_n^l\psi _{S1,h,n}^ -  + c_n^l\psi _{S1,e,n}^ -  + d_n^l\psi _{S1,h,n}^ + ,\\
{\psi _{N,n}} = \sum\limits_{m =  - \infty }^{ + \infty } {\left[ {\left( {{a_m}\psi _{N,e,m}^ +  + {b_m}\psi _{N,h,m}^ -  + {c_m}\psi _{N,e,m}^ -  + {d_m}\psi _{N,h,m}^ + } \right){J_{n - m}}\left( {{{{V_1}} \mathord{\left/
 {\vphantom {{{V_1}} {\hbar \omega }}} \right.
 \kern-\nulldelimiterspace} {\hbar \omega }}} \right)} \right]} ,\\
{\psi _{S2,n}} = a_n^r\psi _{S2,e,n}^ -  + b_n^r\psi _{S2,h,n}^ +  + c_n^r\psi _{S2,e,n}^ +  + d_n^r\psi _{S2,h,n}^ - .
\end{array}
\label{FloquetWaveFunction}
\end{equation}
Here, $a_n^{l/r}$ and $b_n^{l/r}$ are the probability amplitudes of the incoming electron and hole
waves from the left/right superconducting electrode of the $n$th Floquet channel with energy $\varepsilon _n$, respectively, while $c_n^{l/r}$ and $d_n^{l/r}$ are those of the outgoing electron and hole waves.
And we have
\begin{equation}
\begin{array}{l}
\psi _{N,e,n}^ \pm  = \left( {\begin{array}{*{20}{c}}
1\\
0
\end{array}} \right){\left( {{k_{e,n}}} \right)^{ - {1 \mathord{\left/
 {\vphantom {1 2}} \right.
 \kern-\nulldelimiterspace} 2}}}\exp \left( { \pm i{k_{e,n}}x} \right),\\
\psi _{N,h,n}^ \pm  = \left( {\begin{array}{*{20}{c}}
0\\
1
\end{array}} \right){\left( {{k_{h,n}}} \right)^{ - {1 \mathord{\left/
 {\vphantom {1 2}} \right.
 \kern-\nulldelimiterspace} 2}}}\exp \left( { \pm i{k_{h,n}}x} \right),\\
\psi _{S1,e,h;n}^ \pm  = \left( {\begin{array}{*{20}{c}}
{{e^{i{{\eta {1_{e,h;n}}} \mathord{\left/
 {\vphantom {{\eta {1_{e,h;n}}} 2}} \right.
 \kern-\nulldelimiterspace} 2}}}}\\
{{e^{ - i{{\eta {1_{e,h;n}}} \mathord{\left/
 {\vphantom {{\eta {1_{e,h;n}}} 2}} \right.
 \kern-\nulldelimiterspace} 2}}}}
\end{array}} \right){\left( {2{q_{e,h;n}}} \right)^{ - {1 \mathord{\left/
 {\vphantom {1 2}} \right.
 \kern-\nulldelimiterspace} 2}}}{\left( {{{{\varepsilon ^2}} \mathord{\left/
 {\vphantom {{{\varepsilon ^2}} {\Delta _0^2}}} \right.
 \kern-\nulldelimiterspace} {\Delta _0^2}} - 1} \right)^{ - {1 \mathord{\left/
 {\vphantom {1 4}} \right.
 \kern-\nulldelimiterspace} 4}}}\exp \left( { \pm i{q_{e,h;n}}x} \right),
\end{array}
\label{WaveFunctionInSAndNRegions}
\end{equation}
while $\psi _{S2,e,h;n}^ \pm $ has ${\eta {1_{e,h;n}}}$ replaced by ${\eta {2_{e,h;n}}}$ and $\exp \left( { \pm i{q_{e,h;n}}x} \right)$ replaced by $\exp \left[ { \pm i{q_{e,h;n}}\left( {x - L} \right)} \right]$
with ${k_{e,h;n}} = {\left( {{{2m} \mathord{\left/
 {\vphantom {{2m} {{\hbar ^2}}}} \right.
 \kern-\nulldelimiterspace} {{\hbar ^2}}}} \right)^{{1 \mathord{\left/
 {\vphantom {1 2}} \right.
 \kern-\nulldelimiterspace} 2}}}{\left( {{E_F} + {\sigma _{e,h}}{\varepsilon _n}} \right)^{{1 \mathord{\left/
 {\vphantom {1 2}} \right.
 \kern-\nulldelimiterspace} 2}}}$, ${q_{e,h;n}} = {\left( {{{2m} \mathord{\left/
 {\vphantom {{2m} {{\hbar ^2}}}} \right.
 \kern-\nulldelimiterspace} {{\hbar ^2}}}} \right)^{{1 \mathord{\left/
 {\vphantom {1 2}} \right.
 \kern-\nulldelimiterspace} 2}}}{\left[ {{E_F} + {\sigma _{e,h}}{{\left( {\varepsilon _n^2 - \Delta _0^2} \right)}^{{1 \mathord{\left/
 {\vphantom {1 2}} \right.
 \kern-\nulldelimiterspace} 2}}}} \right]^{{1 \mathord{\left/
 {\vphantom {1 2}} \right.
 \kern-\nulldelimiterspace} 2}}}$, $\eta {1_{e,h;n}} = {\phi  \mathord{\left/
 {\vphantom {\phi  2}} \right.
 \kern-\nulldelimiterspace} 2} + {\sigma _{e,h}}\arccos \left( {{{{\varepsilon _n}} \mathord{\left/
 {\vphantom {{{\varepsilon _n}} {{\Delta _0}}}} \right.
 \kern-\nulldelimiterspace} {{\Delta _0}}}} \right)$, and $\eta {2_{e,h;n}} =  - {\phi  \mathord{\left/
 {\vphantom {\phi  2}} \right.
 \kern-\nulldelimiterspace} 2} + {\sigma _{e,h}}\arccos \left( {{{{\varepsilon _n}} \mathord{\left/
 {\vphantom {{{\varepsilon _n}} {{\Delta _0}}}} \right.
 \kern-\nulldelimiterspace} {{\Delta _0}}}} \right)$. $J_n (x)$ are the $n$th-order first kind Bessel functions.

By continuity of $\psi $ and ${{\partial \psi } \mathord{\left/
 {\vphantom {{\partial \psi } {\partial x}}} \right.
 \kern-\nulldelimiterspace} {\partial x}}$ at the normal-superconducting interfaces, the electron-hole Floquet scattering matrix expressed as
\begin{equation}
\left( {\begin{array}{*{20}{c}}
{c_n^l}\\
{d_n^l}\\
{c_n^r}\\
{d_n^r}
\end{array}} \right) = \sum\limits_m {{S_{nm}}\left( {\begin{array}{*{20}{c}}
{a_m^l}\\
{b_m^l}\\
{a_m^r}\\
{b_m^r}
\end{array}} \right)}=\sum\limits_m { \left( {\begin{array}{*{20}{c}}
{{r_{nm}^{ee}}}&{{r_{nm}^{eh}}}&{{t'}_{nm}^{ee}}&{{t'}_{nm}^{eh}}\\
{{r_{nm}^{he}}}&{{r_{nm}^{hh}}}&{{t'}_{nm}^{he}}&{{t'}_{nm}^{hh}}\\
{{t_{nm}^{ee}}}&{{t_{nm}^{eh}}}&{{r'}_{nm}^{ee}}&{{r'}_{nm}^{eh}}\\
{{t_{nm}^{he}}}&{{t_{nm}^{hh}}}&{{r'}_{nm}^{he}}&{{r'}_{nm}^{hh}}
\end{array}} \right) \left( {\begin{array}{*{20}{c}}
{a_m^l}\\
{b_m^l}\\
{a_m^r}\\
{b_m^r}
\end{array}} \right) } ,
\end{equation}
can be obtained by matrix algebra (see
the Appendix).
We define the total electron and hole reflection and transmission coefficients as
\begin{equation}
\begin{array}{*{20}{l}}
{{R_{ee}} = \sum\limits_{n =  - \infty }^{ + \infty } {{{\left| {r_{0n}^{ee}} \right|}^2}} ,}&{{R_{eh}} = \sum\limits_{n =  - \infty }^{ + \infty } {{{\left| {r_{0n}^{eh}} \right|}^2}} ,}\\
{{T_{ee}} = \sum\limits_{n =  - \infty }^{ + \infty } {{{\left| {t_{0n}^{ee}} \right|}^2}} ,}&{{T_{eh}} = \sum\limits_{n =  - \infty }^{ + \infty } {{{\left| {t_{0n}^{eh}} \right|}^2}} .}
\end{array}
\label{TotalReflectionAndTransmisionCoefficients}
\end{equation}
The summation is over all propagating modes and with a cutoff of $n_{\max}$ in numerical treatment. The Wigner-Smith matrix can be obtained by $Q\left( \varepsilon  \right) =  - i\hbar {S^\dag }{{\partial S} \mathord{\left/
 {\vphantom {{\partial S} {\partial \varepsilon }}} \right.
 \kern-\nulldelimiterspace} {\partial \varepsilon }}$. The DOS $\rho \left( \varepsilon  \right)$ is directly related to $Q\left( \varepsilon  \right)$ and the WSDT $\tau _n$ by Eq. (\ref{DOS}). We can also define the total WSDT as ${\tau _{{\rm{WS}}}} = \sum\limits_n {{\tau _n}} $. Biased by a current source flowing from the $S_1$ to the $S_2$ superconducting electrodes and driven by a high-frequency-ac-potential in the normal region, the supercurrent flowing from the $S_1$ to the $S_2$ region can be calculated by the general relation\cite{BeenakkerPRL1991}
\begin{equation}
I = {I_1} + {I_2},
\label{I}
\end{equation}
with
\begin{equation}
{I_1} =  - \frac{{2e}}{\hbar }\tanh \left( {{{{E_b}} \mathord{\left/
 {\vphantom {{{E_b}} {2{k_B}T}}} \right.
 \kern-\nulldelimiterspace} {2{k_B}T}}} \right)\frac{{d{E_b}}}{{d\phi }},
\label{I1}
\end{equation}
\begin{equation}
{I_2} =  - \frac{{2e}}{\hbar }2{k_B}T\int_{{\Delta _0}}^\infty  {d\varepsilon \ln \left[ {2\cosh \left( {{\varepsilon  \mathord{\left/
 {\vphantom {\varepsilon  {2{k_B}T}}} \right.
 \kern-\nulldelimiterspace} {2{k_B}T}}} \right)} \right]\frac{{\partial \rho }}{{\partial \phi }}} .
\label{I2}
\end{equation}

\section{Numerical Results and Interpretations}

In our numerical treatment, experimentally realistic parameters are used. $E_F =20$ meV, $\Delta _0 =2$ to $3$ meV, $\hbar \omega =5$ to $6$ meV, $V_1 =1$ meV, and $n_{\max} =1 > V_1 / \hbar \omega $. Numerical results of the total electron and hole reflection and transmission coefficients defined in Eq. (\ref{TotalReflectionAndTransmisionCoefficients}) are shown in Fig. 3. Sharp resonances occur in $R_{ee}$, $R_{eh}$, and $T_{eh}$; sharp antiresonances occur in $T_{ee}$, when one of the electron or hole Floquet sidebands coincides with the bound quasiparticle state $E_b$ by $\varepsilon _{e} - \hbar \omega  = {E_b}$ in the electron channel and $\hbar \omega  - \varepsilon _{h} = {E_b}$ in the hole channel within an accuracy of $0.001$ meV. $E_b =1.99519 $ and $1.99487$ meV can be obtained by solution of the secular equation for $\phi =0$ and $0.025$ in radian, respectively. Current conservation secures unitarity of the Floquet scattering matrix $S$. It can be seen from Fig. 3 that $R_{ee}+R_{eh}+T_{ee}+T_{eh} =1$, which confirms unitarity of the $S$ matrix. From Fig. 2 it can be seen that $E_b$ decreases as $\phi$ in creases for $\phi < \pi$. As $\phi $ increases from $0$ to $0.025$ in radian, resonance from the electron Floquet channel occurs at a larger energy and resonance from the hole channel occurs at a smaller energy according to the relation $\varepsilon _{e} - \hbar \omega  = {E_b}$ and $\hbar \omega  - \varepsilon _{h} = {E_b}$. It can also be seen that the resonance peaks of electrons are broader than that of the holes. This is because that the virtual mass of the hole is larger than that of the electron and the characteristic time of the former is longer than the latter, which is prominently seen in Fig. 4. In the range $ \Delta _0 < \varepsilon < \hbar \omega + \Delta _0 $ we considered, $k _{e,h;0}$, $k _{e,h;\pm 1}$, $q _{e,h;0}$, and $q _{e,h; \pm 1}$ are all real in the normal region; $k _{e,h;0}$, $k _{e,h; + 1}$, $q _{e,h;0}$, and $q _{e,h; + 1}$ are real in the superconducting regions; $k _{e,h; - 1}$ is imaginary and $q _{e,h; - 1}$ complex in the two superconducting regions evanescent to $\pm \infty$. The incident quasiparticle state transfers through the propagating quasiparticle Floquet sideband and interferes with the direct transmission giving rise to a resonance.

Sharp resonances also occur in the $\tau _{\rm{WS}}$ (see Fig. 4) featuring the characteristics of the Floquet scattering matrix. The resonant $\tau _{\rm{WS}}$ of the hole is larger than that of the hole because of larger virtual hole mass, which is natural in electron-hole systems. Our numerical results also show that $\tau _{\rm{WS}}$ outside of the resonant peak is $10^{-14} $ to $10^{-13}$ s, a normal WSDT in nanoscale tunneling. Values of the $\tau _{\rm{WS}}$ at resonance are as high as $10^{-7}$ s, which is a dramatic enhancement in comparison with the normal $10^{-14} $ s and possible to be observed in experiment.

The energy gap of BCS superconductors $\Delta _0$ is approximately related to the temperature by ${\Delta _0} = 3.2{k_B}{T_c}{\left( {1 - {T \mathord{\left/
 {\vphantom {T {{T_c}}}} \right.
 \kern-\nulldelimiterspace} {{T_c}}}} \right)^{{1 \mathord{\left/
 {\vphantom {1 2}} \right.
 \kern-\nulldelimiterspace} 2}}}$ with $T_c$ the critical temperature\cite{BardeenPR1957}. From Fig. 2 it can be seen that the bound quasiparticle energy increases linearly with $\Delta _0$ for fixed $E_F$, $L$, and $\phi $. It can also be seen from Figs. 3 and 4 that resonances in the reflection and transmission coefficients and $\tau _{\rm{WS}}$ have extremely sharp peaks with their widths smaller than $0.0002$ meV in a wide energy window of $5$ meV. The peak energy is $\varepsilon _{e} - \hbar \omega  = {E_b}$ and $\hbar \omega  - \varepsilon _{h} = {E_b}$ within an accuracy of $0.001$ meV. These results based on the time-dependent SNS Josephson junction suggest a potential way to detect the bound quasiparticle state energy, the superconducting energy gap, and hence the superconducting transition temperature.

Supercurrent of the considered device are calculated by Eqs. (\ref{I}) to (\ref{I2}). The numerical results are given in Fig. 5. The supercurrent flowing through a static Josephson SNS junction is\cite{BardeenPRB1972} in the order of ${\rm{meV}} \cdot {e \mathord{\left/
 {\vphantom {e \hbar }} \right.
 \kern-\nulldelimiterspace} \hbar }$. It can be seen from Fig. 5 that the supercurrent is dramatically enhanced in magnitude by the nonadiabatic driving force. This can be interpreted by the change in the continuous DOS above $\Delta _0$ as a result of the Floquet states formed. As demonstrated in Eq. (\ref{DOS}), the DOS is proportional to the sum of the eigenvalues of the Wigner-Smith matrix $Q\left( \varepsilon  \right) =  - i\hbar {S^\dag }{{\partial S} \mathord{\left/
 {\vphantom {{\partial S} {\partial \varepsilon }}} \right.
 \kern-\nulldelimiterspace} {\partial \varepsilon }}$ with $S$ the Floquet scattering matrix. The eigenvalues of the Wigner-Smith matrix corresponds to the WSDT. The Floquet sidebands are formed in the high-frequency driven transport processes. The sidebands contribute to the dynamic DOS and enhance the supercurrent. From Figs. 3 and 4, it can be seen that when one of the Floquet sidebands coincides with the bound quasiparticle state, sharp resonances occur in the transmission and WSDT spectrum, which further enhances the supercurrent. Enhancement by the resonances is vital as the direction of the supercurrent is reversed against the original current bias. Energies of the peaks in $\tau _{\rm{WS}}$ and hence in $\rho $ corresponding to the hole-Floquet-channel resonance decreases with $\phi $ while energies of the peaks corresponding to the electron-Floquet-channel resonance increases with $\phi$ governed by the relations $\varepsilon _{e} - \hbar \omega  = {E_b}$ and $\hbar \omega  - \varepsilon _{h} = {E_b}$ as $E_b$ decreases with $\phi $ for $\phi < \pi$. $I_1$ in Eq. (\ref{I1}) is proportional to $ - {{d{E_b}} \mathord{\left/
 {\vphantom {{d{E_b}} {d\phi }}} \right.
 \kern-\nulldelimiterspace} {d\phi }}$ and is positive for $\phi < \pi$ and negative for $\phi > \pi$. $I_2$ in Eq. (\ref{I2}) is proportional to the integral of $ - {{\partial \rho } \mathord{\left/
 {\vphantom {{\partial \rho } {\partial \phi }}} \right.
 \kern-\nulldelimiterspace} {\partial \phi }}$ over $\varepsilon $. Contribution of the hole-channel peak in $\rho $ to $I_2$ is negative for $\phi < \pi$ and positive for $\phi > \pi$, while contribution of the electron-channel peak in $\rho $ to $I_2$ is reversed. Since the hole-channel peak in $\rho$ is higher and sharper than the electron-channel, contribution of the former to $I_2$ overweighs that of the latter, giving rise to the sign reversal of the supercurrent. As the resonances are extremely strong, the supercurrent in the nonadiabatic process is several orders larger than the static supercurrent. Since $ - {{\partial \rho } \mathord{\left/
 {\vphantom {{\partial \rho } {\partial \phi }}} \right.
 \kern-\nulldelimiterspace} {\partial \phi }}$ is nonzero even when $\rho $ is small at the energies away from the resonant peaks, the supercurrent varies in an irregular pattern as it is an integral result of $ - {{\partial \rho } \mathord{\left/
 {\vphantom {{\partial \rho } {\partial \phi }}} \right.
 \kern-\nulldelimiterspace} {\partial \phi }}$. For $\phi > \pi$, $E_b$ increases with $\phi$. The supercurrent is positive or negatively small as a combined result of the resonant peaks and the integral of $ - {{\partial \rho } \mathord{\left/
 {\vphantom {{\partial \rho } {\partial \phi }}} \right.
 \kern-\nulldelimiterspace} {\partial \phi }}$ over the large range of energy. The resonant peaks are sharper for stronger driving forces. As a result the supercurrent for $\hbar \omega =6$ meV is larger than $\hbar \omega =5$ meV for $\phi < \pi$.

\section{Conclusions}

In conclusion, resonances are observed in the nonadiabatic transmission and WSDT spectra of the time-dependent Josephson SNS junction as a result of quantum path interferences. In the picture of Floquet scattering, the electron and hole Floquet sidebands are formed. When one of the electron or hole Floquet sidebands coincides with the bound quasiparticle state within the energy gap of the superconductors at $\varepsilon _{e} - \hbar \omega  = {E_b}$ and $\hbar \omega  - \varepsilon _{h} = {E_b}$, resonances occur in $R_{ee}$, $R_{eh}$, $T_{eh}$, and $\tau _{\rm{WS}}$; antiresonances occur in $T_{ee}$. The resonances are extremely sharp suggesting a potential experimental determination of $E_b$. Using the Wigner-Smith matrix, the dynamic DOS and supercurrent biased by a current source are obtained. Sharp resonances in the dynamic DOS consisting of Floquet modes greatly enhance the supercurrent in comparison with the condition without a pumping force.

\section{Acknowledgements}

This project was supported by the National Natural Science
Foundation of China (No. 11004063) and the Fundamental Research
Funds for the Central Universities, SCUT (No. 2014ZG0044).

\section{Appendix: Derivation of the Floquet Scattering Matrix }

Continuity equations of $\psi $ and ${{\partial \psi } \mathord{\left/
 {\vphantom {{\partial \psi } {\partial x}}} \right.
 \kern-\nulldelimiterspace} {\partial x}}$ [The Floquet state wave functions are defined in Eqs. (\ref{TimeDependentWaveFunction}) to (\ref{WaveFunctionInSAndNRegions})] at the normal-superconducting interfaces $x=0$ and $x=L$ can be expressed in matrix form as follows.
\begin{equation}
M_{1u}^e{a^l} + M_{1u}^h{b^l} + M_{1u}^e{c^l} + M_{1u}^h{d^l} = M_J^ea + M_J^ec,
\label{One}
\end{equation}
\begin{equation}
M_{1d}^e{a^l} + M_{1d}^h{b^l} + M_{1d}^e{c^l} + M_{1d}^h{d^l} = M_J^hb + M_J^hd,
\end{equation}
\begin{equation}
M_{2 + }^ea + M_{2 - }^ec = M_{2u}^e{a^r} + M_{2u}^h{b^r} + M_{2u}^e{c^r} + M_{2u}^h{d^r},
\end{equation}
\begin{equation}
M_{2 - }^hb + M_{2 + }^hd = M_{2d}^e{a^r} + M_{2d}^h{b^r} + M_{2d}^e{c^r} + M_{2d}^h{d^r},
\end{equation}
\begin{equation}
M_{4u1 + }^e{a^l} + M_{4u1 - }^h{b^l} + M_{4u1 - }^e{c^l} + M_{4u1 + }^h{d^l} = M_{5 + }^ea + M_{5 - }^ec,
\end{equation}
\begin{equation}
M_{4d1 + }^e{a^l} + M_{4d1 - }^h{b^l} + M_{4d1 - }^e{c^l} + M_{4d1 + }^h{d^l} = M_{5 - }^hb + M_{5 + }^hd,
\end{equation}
\begin{equation}
M_{8 + }^ea + M_{8 - }^ec = M_{4u2 - }^e{a^r} + M_{4u2 + }^h{b^r} + M_{4u2 + }^e{c^r} + M_{4u2 - }^h{d^r},
\end{equation}
\begin{equation}
M_{8 - }^hb + M_{8 + }^hd = M_{4d2 - }^e{a^r} + M_{4d2 + }^h{b^r} + M_{4d2 + }^e{c^r} + M_{4d2 - }^h{d^r}.
\label{Eight}
\end{equation}
$a^{l/r}$, $b^{l/r}$, $c^{l/r}$, and $d^{l/r}$ are column vectors made up of elements $a_n^{l/r}$, $b_n^{l/r}$, $c_n^{l/r}$, and $d_n^{l/r}$, respectively. Elements of the coefficient matrices are defined as follows:
\begin{equation}
{\left( {M_{1u}^{e,h}} \right)_{nm}} = \exp \left( {i{{\eta {1_{e,h;n}}} \mathord{\left/
 {\vphantom {{\eta {1_{e,h;n}}} 2}} \right.
 \kern-\nulldelimiterspace} 2}} \right){\left( {2{q_{e,h;n}}} \right)^{ - {1 \mathord{\left/
 {\vphantom {1 2}} \right.
 \kern-\nulldelimiterspace} 2}}}{\left( {{{\varepsilon _n^2} \mathord{\left/
 {\vphantom {{\varepsilon _n^2} {\Delta _0^2}}} \right.
 \kern-\nulldelimiterspace} {\Delta _0^2}} - 1} \right)^{ - {1 \mathord{\left/
 {\vphantom {1 4}} \right.
 \kern-\nulldelimiterspace} 4}}}{\delta _{n,m}},
\end{equation}
\begin{equation}
{\left( {M_{1d}^{e,h}} \right)_{nm}} = \exp \left( { - i{{\eta {1_{e,h;n}}} \mathord{\left/
 {\vphantom {{\eta {1_{e,h;n}}} 2}} \right.
 \kern-\nulldelimiterspace} 2}} \right){\left( {2{q_{e,h;n}}} \right)^{ - {1 \mathord{\left/
 {\vphantom {1 2}} \right.
 \kern-\nulldelimiterspace} 2}}}{\left( {{{\varepsilon _n^2} \mathord{\left/
 {\vphantom {{\varepsilon _n^2} {\Delta _0^2}}} \right.
 \kern-\nulldelimiterspace} {\Delta _0^2}} - 1} \right)^{ - {1 \mathord{\left/
 {\vphantom {1 4}} \right.
 \kern-\nulldelimiterspace} 4}}}{\delta _{n,m}},
\end{equation}
\begin{equation}
{\left( {M_J^{e,h}} \right)_{nm}} = {J_{n - m}}\left( {{{{V_1}} \mathord{\left/
 {\vphantom {{{V_1}} {\hbar \omega }}} \right.
 \kern-\nulldelimiterspace} {\hbar \omega }}} \right){\left( {{k_{e,h;m}}} \right)^{ - {1 \mathord{\left/
 {\vphantom {1 2}} \right.
 \kern-\nulldelimiterspace} 2}}},
\end{equation}
\begin{equation}
{\left( {M_{2u}^{e,h}} \right)_{nm}} = \exp \left( {{{i\eta {2_{e,h;n}}} \mathord{\left/
 {\vphantom {{i\eta {2_{e,h;n}}} 2}} \right.
 \kern-\nulldelimiterspace} 2}} \right){\left( {2{q_{e,h;n}}} \right)^{ - {1 \mathord{\left/
 {\vphantom {1 2}} \right.
 \kern-\nulldelimiterspace} 2}}}{\left( {{{\varepsilon _n^2} \mathord{\left/
 {\vphantom {{\varepsilon _n^2} {\Delta _0^2}}} \right.
 \kern-\nulldelimiterspace} {\Delta _0^2}} - 1} \right)^{ - {1 \mathord{\left/
 {\vphantom {1 4}} \right.
 \kern-\nulldelimiterspace} 4}}}{\delta _{n,m}},
\end{equation}
\begin{equation}
{\left( {M_{2d}^{e,h}} \right)_{nm}} = \exp \left( { - {{i\eta {2_{e,h;n}}} \mathord{\left/
 {\vphantom {{i\eta {2_{e,h;n}}} 2}} \right.
 \kern-\nulldelimiterspace} 2}} \right){\left( {2q_{sn}^{e,h}} \right)^{ - {1 \mathord{\left/
 {\vphantom {1 2}} \right.
 \kern-\nulldelimiterspace} 2}}}{\left( {{{\varepsilon _n^2} \mathord{\left/
 {\vphantom {{\varepsilon _n^2} {\Delta _0^2}}} \right.
 \kern-\nulldelimiterspace} {\Delta _0^2}} - 1} \right)^{ - {1 \mathord{\left/
 {\vphantom {1 4}} \right.
 \kern-\nulldelimiterspace} 4}}}{\delta _{n,m}},
\end{equation}
\begin{equation}
{\left( {M_{2 \pm }^{e,h}} \right)_{nm}} = \exp \left( { \pm i{k_{e,h;m}}L} \right){\left( {{k_{e,h;m}}} \right)^{ - {1 \mathord{\left/
 {\vphantom {1 2}} \right.
 \kern-\nulldelimiterspace} 2}}}{J_{n - m}}\left( {{{{V_1}} \mathord{\left/
 {\vphantom {{{V_1}} {\hbar \omega }}} \right.
 \kern-\nulldelimiterspace} {\hbar \omega }}} \right),
\end{equation}
\begin{equation}
{\left( {M_{4u1 \pm }^{e,h}} \right)_{nm}} =  \pm i{q_{e,h;n}}\exp \left( {i{{\eta {1_{e,h;n}}} \mathord{\left/
 {\vphantom {{\eta {1_{e,h;n}}} 2}} \right.
 \kern-\nulldelimiterspace} 2}} \right){\left( {2{q_{e,h;n}}} \right)^{ - {1 \mathord{\left/
 {\vphantom {1 2}} \right.
 \kern-\nulldelimiterspace} 2}}}{\left( {{{\varepsilon _n^2} \mathord{\left/
 {\vphantom {{\varepsilon _n^2} {\Delta _0^2}}} \right.
 \kern-\nulldelimiterspace} {\Delta _0^2}} - 1} \right)^{ - {1 \mathord{\left/
 {\vphantom {1 4}} \right.
 \kern-\nulldelimiterspace} 4}}}{\delta _{n,m}},
\end{equation}
\begin{equation}
{\left( {M_{4d1 \pm }^{e,h}} \right)_{nm}} =  \pm i{q_{e,h;n}}\exp \left( { - i{{\eta {1_{e,h;n}}} \mathord{\left/
 {\vphantom {{\eta {1_{e,h;n}}} 2}} \right.
 \kern-\nulldelimiterspace} 2}} \right){\left( {2{q_{e,h;n}}} \right)^{ - {1 \mathord{\left/
 {\vphantom {1 2}} \right.
 \kern-\nulldelimiterspace} 2}}}{\left( {{{\varepsilon _n^2} \mathord{\left/
 {\vphantom {{\varepsilon _n^2} {\Delta _0^2}}} \right.
 \kern-\nulldelimiterspace} {\Delta _0^2}} - 1} \right)^{ - {1 \mathord{\left/
 {\vphantom {1 4}} \right.
 \kern-\nulldelimiterspace} 4}}}{\delta _{n,m}},
\end{equation}
\begin{equation}
{\left( {M_{5 \pm }^{e,h}} \right)_{nm}} =  \pm i{k_{e,h;m}}{\left( {{k_{e,h;m}}} \right)^{ - {1 \mathord{\left/
 {\vphantom {1 2}} \right.
 \kern-\nulldelimiterspace} 2}}}{J_{n - m}}\left( {{{{V_1}} \mathord{\left/
 {\vphantom {{{V_1}} {\hbar \omega }}} \right.
 \kern-\nulldelimiterspace} {\hbar \omega }}} \right),
\end{equation}
\begin{equation}
{\left( {M_{4u2 \pm }^{e,h}} \right)_{nm}} =  \pm i{q_{e,h;n}}\exp \left( {i{{\eta {2_{e,h;n}}} \mathord{\left/
 {\vphantom {{\eta {2_{e,h;n}}} 2}} \right.
 \kern-\nulldelimiterspace} 2}} \right){\left( {2{q_{e,h;n}}} \right)^{ - {1 \mathord{\left/
 {\vphantom {1 2}} \right.
 \kern-\nulldelimiterspace} 2}}}{\left( {{{\varepsilon _n^2} \mathord{\left/
 {\vphantom {{\varepsilon _n^2} {\Delta _0^2}}} \right.
 \kern-\nulldelimiterspace} {\Delta _0^2}} - 1} \right)^{ - {1 \mathord{\left/
 {\vphantom {1 4}} \right.
 \kern-\nulldelimiterspace} 4}}}{\delta _{n,m}},
\end{equation}
\begin{equation}
{\left( {M_{4d2 \pm }^{e,h}} \right)_{nm}} =  \pm i{q_{e,h;n}}\exp \left( { - i{{\eta {2_{e,h;n}}} \mathord{\left/
 {\vphantom {{\eta {2_{e,h;n}}} 2}} \right.
 \kern-\nulldelimiterspace} 2}} \right){\left( {2{q_{e,h;n}}} \right)^{ - {1 \mathord{\left/
 {\vphantom {1 2}} \right.
 \kern-\nulldelimiterspace} 2}}}{\left( {{{\varepsilon _n^2} \mathord{\left/
 {\vphantom {{\varepsilon _n^2} {\Delta _0^2}}} \right.
 \kern-\nulldelimiterspace} {\Delta _0^2}} - 1} \right)^{ - {1 \mathord{\left/
 {\vphantom {1 4}} \right.
 \kern-\nulldelimiterspace} 4}}}{\delta _{n,m}},
\end{equation}
\begin{equation}
{\left( {M_{8 \pm }^{e,h}} \right)_{nm}} =  \pm i{k_{e,h;m}}\exp \left( { \pm i{k_{e,h;m}}L} \right){\left( {{k_{e,h;m}}} \right)^{ - {1 \mathord{\left/
 {\vphantom {1 2}} \right.
 \kern-\nulldelimiterspace} 2}}}{J_{n - m}}\left( {{{{V_1}} \mathord{\left/
 {\vphantom {{{V_1}} {\hbar \omega }}} \right.
 \kern-\nulldelimiterspace} {\hbar \omega }}} \right).
\end{equation}
The matrix equations (\ref{One}) to (\ref{Eight}) can be transformed into larger matrix equations as
\begin{equation}
\underbrace {\left( {\begin{array}{*{20}{c}}
{M_{1u}^e}&{M_{1u}^h}&{M_{1u}^e}&{M_{1u}^h}\\
{M_{1d}^e}&{M_{1d}^h}&{M_{1d}^e}&{M_{1d}^h}\\
{M_{4u1 + }^e}&{M_{4u1 - }^h}&{M_{4u1 - }^e}&{M_{4u1 + }^h}\\
{M_{4d1 + }^e}&{M_{4d1 - }^h}&{M_{4d1 - }^e}&{M_{4d1 + }^h}
\end{array}} \right)}_{{Ml}}\left( {\begin{array}{*{20}{c}}
{{a^l}}\\
{{b^l}}\\
{{c^l}}\\
{{d^l}}
\end{array}} \right) = \underbrace {\left( {\begin{array}{*{20}{c}}
{{M_J}}&0&{{M_J}}&0\\
0&{{M_J}}&0&{{M_J}}\\
{M_{5 + }^e}&0&{M_{5 - }^e}&0\\
0&{M_{5 - }^h}&0&{M_{5 + }^h}
\end{array}} \right)}_{{M01}}\left( {\begin{array}{*{20}{c}}
a\\
b\\
c\\
d
\end{array}} \right),
\end{equation}
and
\begin{equation}
\underbrace {\left( {\begin{array}{*{20}{c}}
{M_{2 + }^e}&0&{M_{2 - }^e}&0\\
0&{M_{2 - }^h}&0&{M_{2 + }^h}\\
{M_{8 + }^e}&0&{M_{8 - }^e}&0\\
0&{M_{8 - }^h}&0&{M_{8 + }^h}
\end{array}} \right)}_{M02}\left( {\begin{array}{*{20}{c}}
a\\
b\\
c\\
d
\end{array}} \right) = \underbrace {\left( {\begin{array}{*{20}{c}}
{M_{1u}^e}&{M_{1u}^h}&{M_{1u}^e}&{M_{1u}^h}\\
{M_{1d}^e}&{M_{1d}^h}&{M_{1d}^e}&{M_{1d}^h}\\
{M_{4u2 - }^e}&{M_{4u2 + }^h}&{M_{4u2 + }^e}&{M_{4u2 - }^h}\\
{M_{4d2 - }^e}&{M_{4d2 + }^h}&{M_{4d2 + }^e}&{M_{4d2 - }^h}
\end{array}} \right)}_{Mr}\left( {\begin{array}{*{20}{c}}
{{a^r}}\\
{{b^r}}\\
{{c^r}}\\
{{d^r}}
\end{array}} \right).
\end{equation}
Then we have
\begin{equation}
\left( {\begin{array}{*{20}{c}}
{{a^r}}\\
{{b^r}}\\
{{c^r}}\\
{{d^r}}
\end{array}} \right) = {\left( {Mr} \right)^{ - 1}} \cdot \left( {M02} \right) \cdot {\left( {M01} \right)^{ - 1}} \cdot \left( {Ml} \right) \cdot \left( {\begin{array}{*{20}{c}}
{{a^l}}\\
{{b^l}}\\
{{c^l}}\\
{{d^l}}
\end{array}} \right).
\end{equation}
By defining
\begin{equation}
M_{ltor} = {\left( {Mr} \right)^{ - 1}} \cdot \left( {M02} \right) \cdot {\left( {M01} \right)^{ - 1}} \cdot \left( {Ml} \right),
\end{equation}
\begin{equation}
{M_{ltor}} = \left( {\begin{array}{*{20}{c}}
{{W_{aa}}}&{{W_{ab}}}&{{W_{ac}}}&{{W_{ad}}}\\
{{W_{ba}}}&{{W_{bb}}}&{{W_{bc}}}&{{W_{bd}}}\\
{{W_{ca}}}&{{W_{cb}}}&{{W_{cc}}}&{{W_{cd}}}\\
{{W_{da}}}&{{W_{db}}}&{{W_{dc}}}&{{W_{dd}}}
\end{array}} \right),
\end{equation}
\begin{equation}
{M_{cd}} = \left( {\begin{array}{*{20}{c}}
{ - {W_{ac}}}&{ - {W_{ad}}}&0&0\\
{ - {W_{bc}}}&{ - {W_{bd}}}&0&0\\
{ - {W_{cc}}}&{ - {W_{cd}}}&1&0\\
{ - {W_{dc}}}&{ - {W_{dd}}}&0&1
\end{array}} \right),
\label{Mcd}
\end{equation}
\begin{equation}
{M_{ab}} = \left( {\begin{array}{*{20}{c}}
{{W_{aa}}}&{{W_{ab}}}&{ - 1}&0\\
{{W_{ba}}}&{{W_{bb}}}&0&{ - 1}\\
{{W_{ca}}}&{{W_{cb}}}&0&0\\
{{W_{da}}}&{{W_{db}}}&0&0
\end{array}} \right),
\label{Mab}
\end{equation}
we can obtain the electron-hole Floquet scattering matrix as
\begin{equation}
S = {\left( {{M_{cd}}} \right)^{ - 1}} \cdot {M_{ab}}.
\end{equation}
In Eqs. (\ref{Mcd}) and (\ref{Mab}), ``$0$" is a zero $\left( {2{n_{{\rm{max}}}} + 1} \right) \times \left( {2{n_{{\rm{max}}}} + 1} \right)$ matrix and ``$1$" is a unitary $\left( {2{n_{{\rm{max}}}} + 1} \right) \times \left( {2{n_{{\rm{max}}}} + 1} \right)$ matrix with $n_{{\rm{max}}}$ the maximal Floquet channel index.

\clearpage

\begin{figure}[h]
\includegraphics[height=10cm, width=14cm]{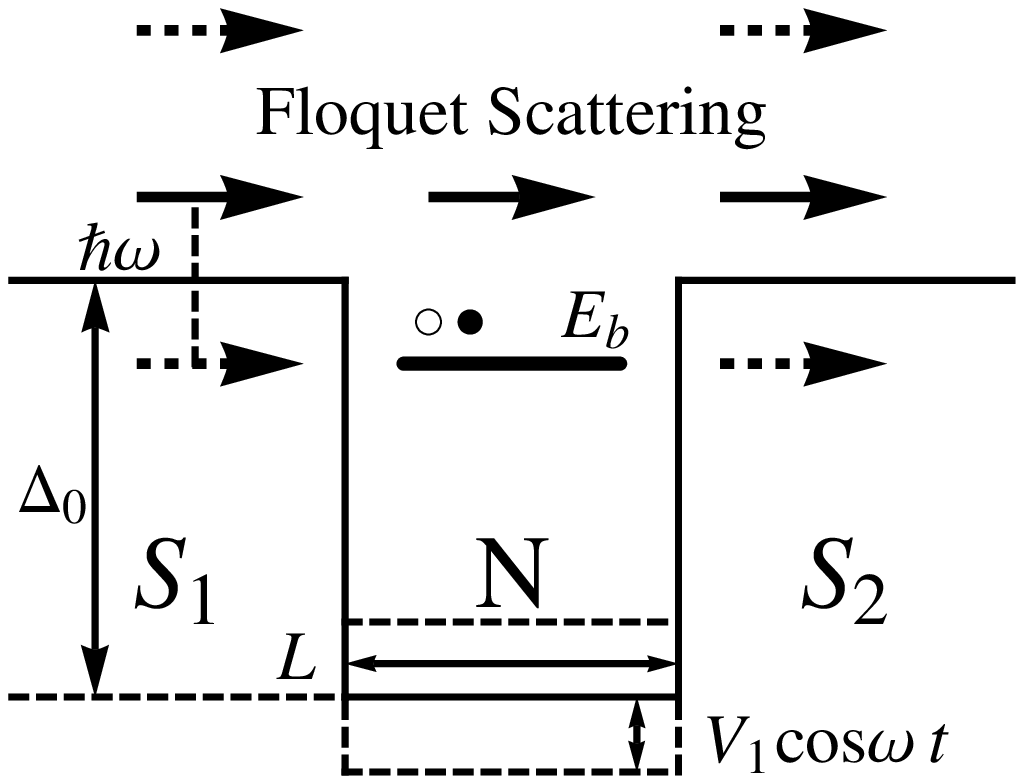}
\caption{Model for the one-dimensional time-dependent Josephson SNS junction. Width of the normal region is $L$. Order parameters of $S_1$ and $S_2$ regions have the same absolute value of $\Delta _0$ and phase difference of $\phi $. When $\Delta _0$ and the Fermi energy $E_F$ is small, only one bound quasiparticle state with the energy of $E_b$ is confined in the normal region consisting of equal probabilities of particle and hole states. A high-frequency-ac-gate potential of $V_1 \cos \omega t$ is applied to the normal region. Driven by the time-dependent potential, Floquet sidebands are formed with energy spacing $\hbar \omega$. When the continuous spectrum above the energy gap has the electron or hole Floquet sideband falling within the energy gap and coincident with the bound state, resonance or antiresonance occurs in the transmission coefficients and the Wigner-Smith delay times.     }
\end{figure}

\clearpage

\begin{figure}[h]
\includegraphics[height=10cm, width=14cm]{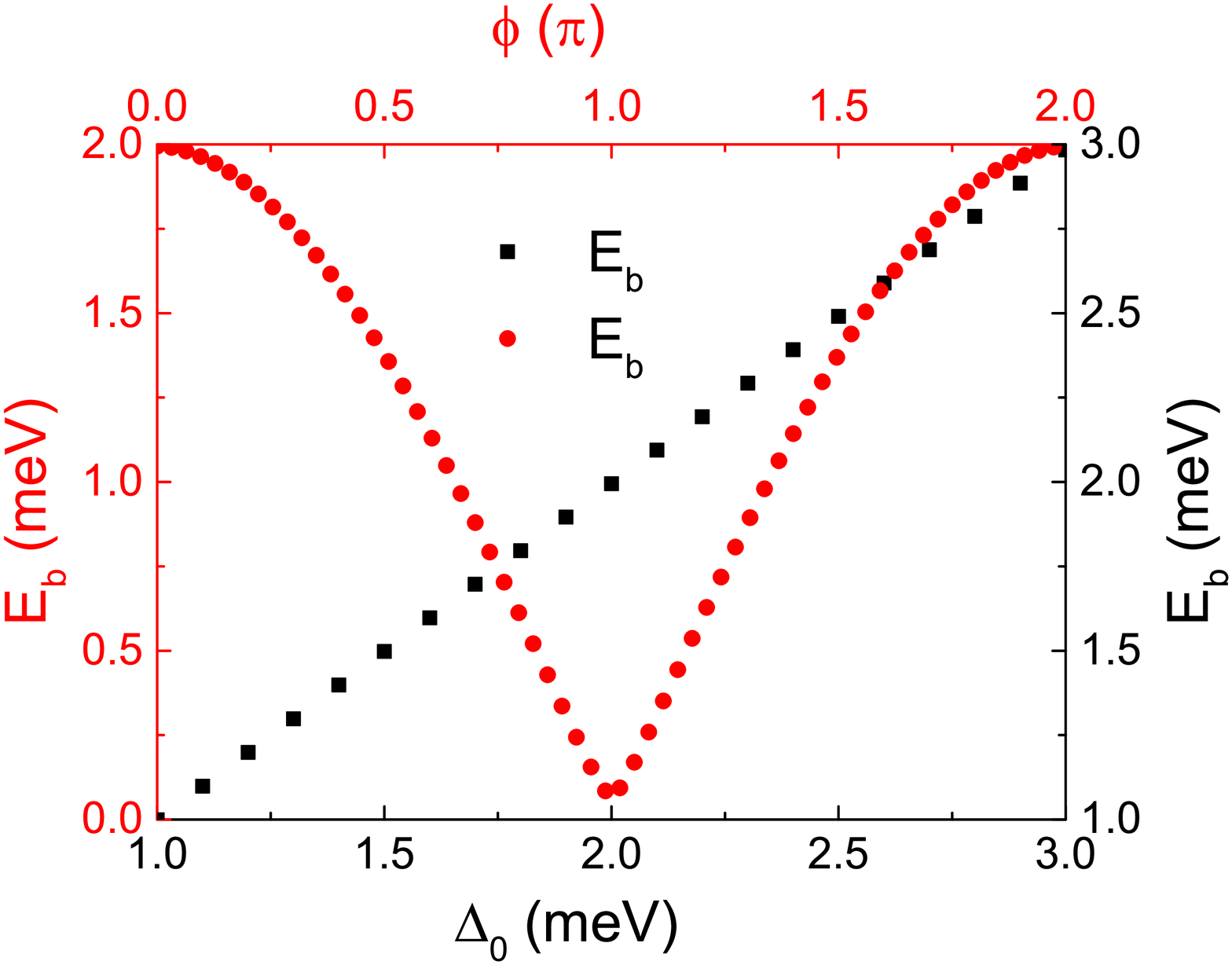}
\caption{Bound quasiparticle energy $E_b$ measured from the Fermi energy $E_F$. Its variation as functions of $\Delta _0$ and $\phi$ corresponds to black and red symbols, respectively. Corresponding coordinates are in the same colors as the symbols. The parameters are $E_F=20$ meV, $L=10$ \AA, $\phi=0$ for variation as a function of $\Delta _0$ and $\Delta _0$=2 meV for variation as a function of $\phi$, respectively.  }
\end{figure}

\begin{figure}[h]
\includegraphics[height=10cm, width=15cm]{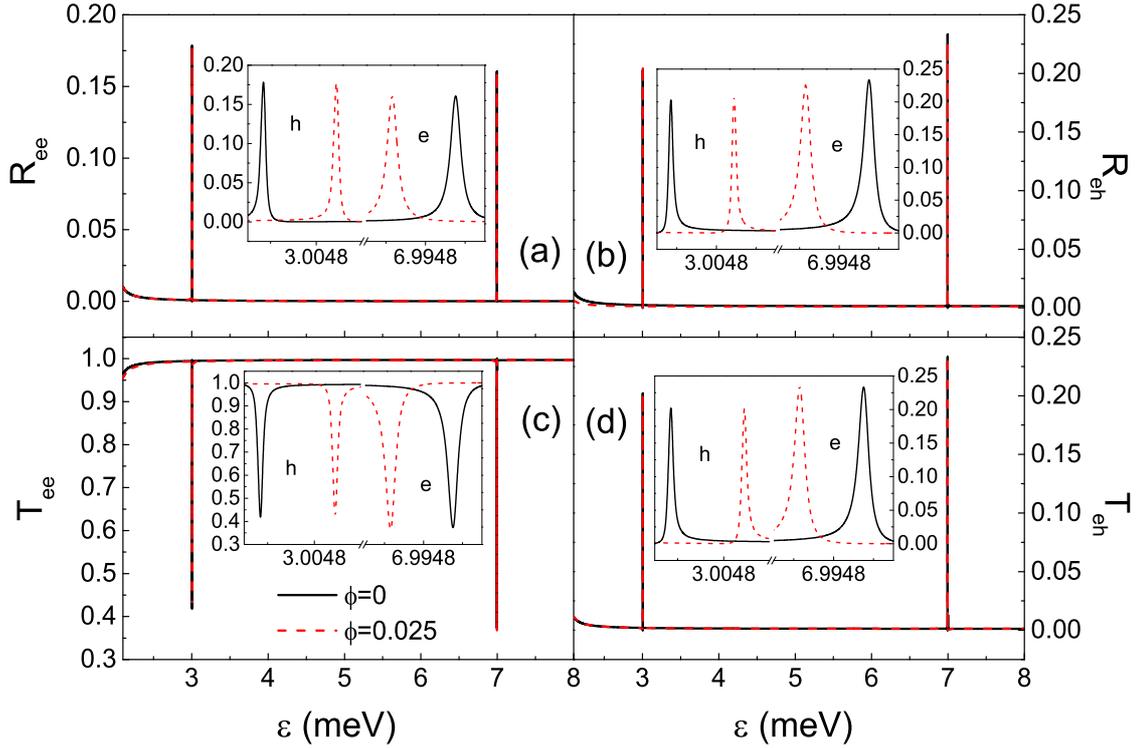}
\caption{ Numerical results of the total reflection and transmission coefficients (a) $R_{ee}$, (b) $R_{eh}$, (c) $T_{ee}$, and (d) $T_{eh}$ as a function of the incident quasiparticle energy $\varepsilon $ for different $\phi $. Sharp resonances occur in $R_{ee}$, $R_{eh}$, and $T_{eh}$ and sharp antiresonances occur in $T_{ee}$, when one of the electron or hole Floquet sidebands coincides with the bound quasiparticle state $E_b$ by $\varepsilon _{e} - \hbar \omega  = {E_b}$ in the electron channel and $\hbar \omega  - \varepsilon _{h} = {E_b}$ in the hole channel within an accuracy of $0.001$ meV. Insets are zoom-in of the two resonant peaks. The parameters are $E_F=20$ meV, $L=10$ \AA, $\Delta _0 =2$ meV, $\hbar \omega =5$ meV, $V_1 =1$ meV, and $n_{\max} =1$. $E_b =1.99519 $ and $1.99487$ meV for $\phi =0$ and $0.025$ in radian, respectively.  }
\end{figure}

\begin{figure}[h]
\includegraphics[height=10cm, width=13cm]{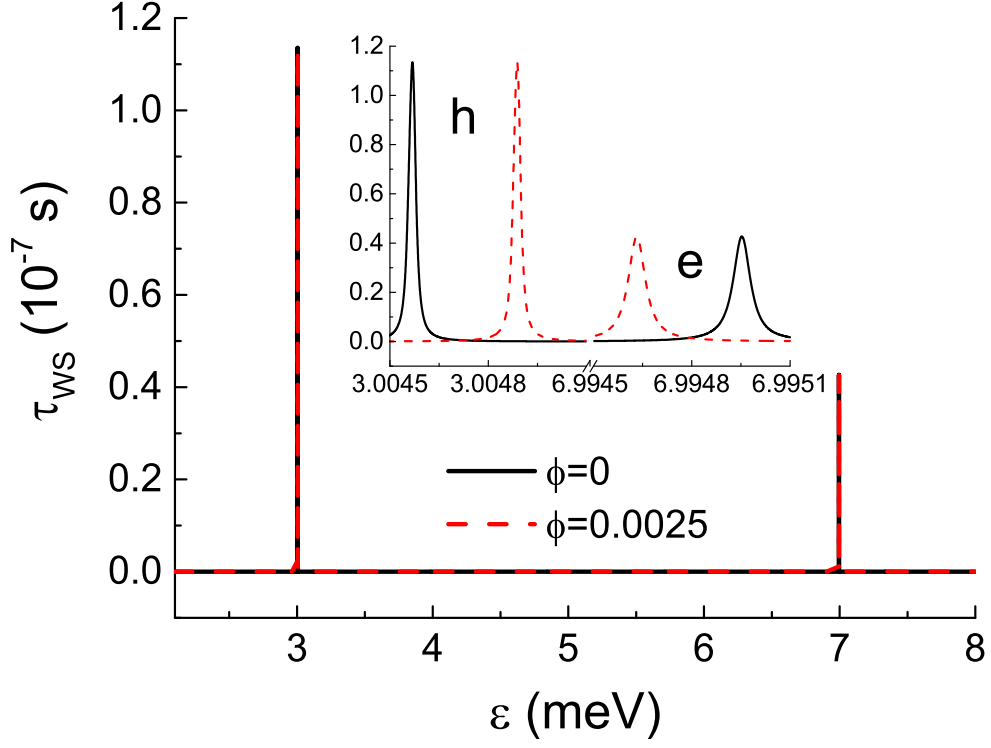}
\caption{ Numerical results of the total WSDT $\tau _{\rm{WS}}$ as a function of the incident quasiparticle energy $\varepsilon $ for different $\phi $. Sharp resonances occur when one of the electron or hole Floquet sidebands coincides with the bound quasiparticle state $E_b$ by $\varepsilon _{e} - \hbar \omega  = {E_b}$ in the electron channel and $\hbar \omega  - \varepsilon _{h} = {E_b}$ in the hole channel within an accuracy of $0.001$ meV. Insets are zoom-in of the two resonant peaks. The parameters are the same as Fig. 3. }
\end{figure}

\begin{figure}[h]
\includegraphics[height=10cm, width=13cm]{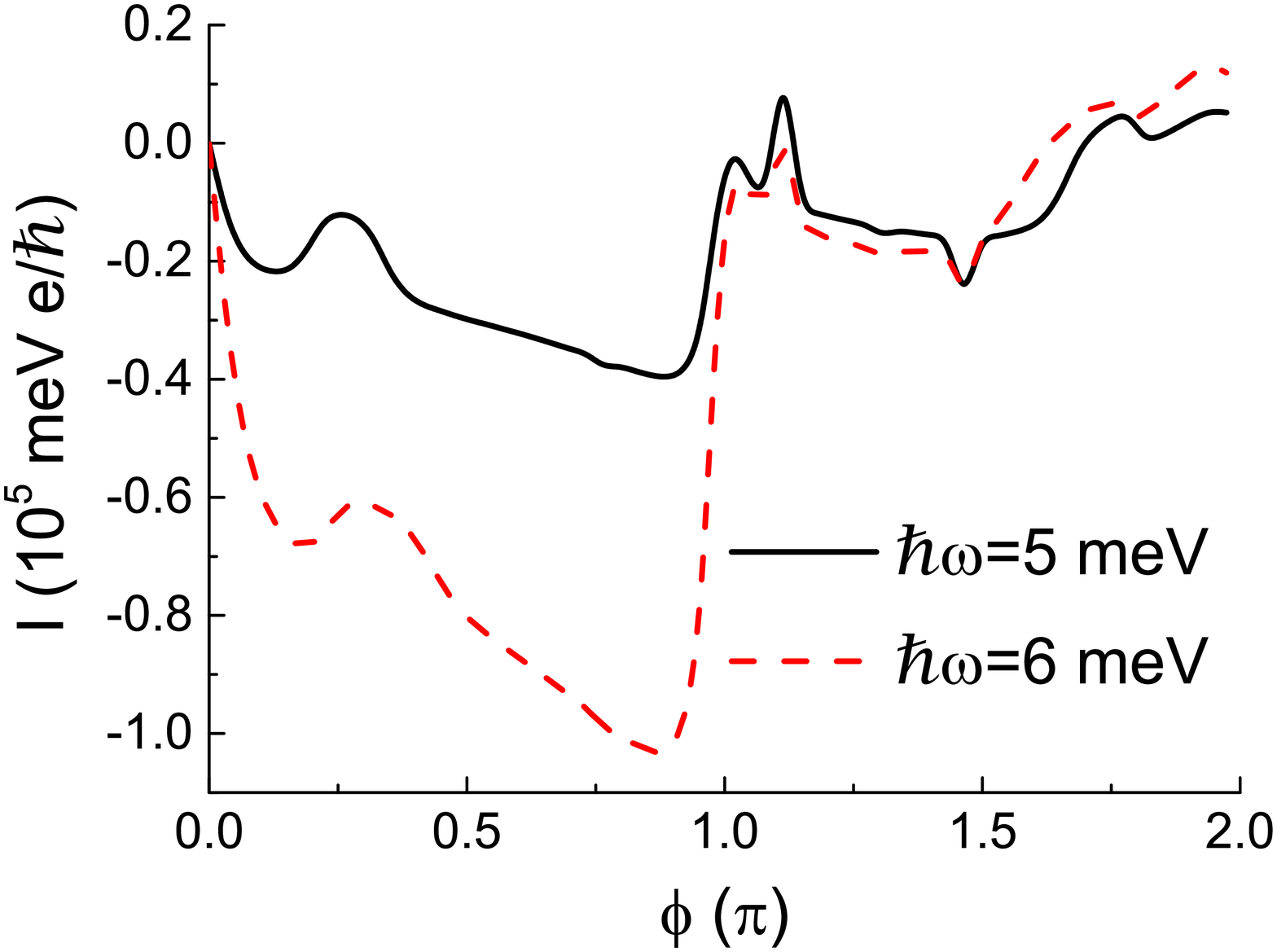}
\caption{ Supercurrent as a function of $\phi $ for different driving frequencies. The parameters are $E_F=20$ meV, $L=10$ \AA, $\Delta _0 =2$ meV, $V_1 =1$ meV, and $n_{\max} =1$. We assume $T_c=30$ K and hence $T=28.24$ K.  }
\end{figure}

\end{document}